# A Network Resource Allocation Recommendation Method with An Improved Similarity Measure

Huiyu Li, Pei Liang, Junhua Hu*

School of Business, Central South University, Changsha 410083, China

Email address: huiyu_li@csu.edu.cn (H.L.); shirley_lp@csu.edu.cn (P.L.) hujunhua@csu.edu.cn (J.H.);

Corresponding author*: Junhua Hu (Email: hujunhua@csu.edu.cn)

**Abstract**

Recommender systems have been acknowledged as efficacious tools for managing information overload. Nevertheless, conventional algorithms adopted in such systems primarily emphasize precise recommendations and, consequently, overlook other vital aspects like the coverage, diversity, and novelty of items. This approach results in less exposure for long-tail items. In this paper, to personalize the recommendations and allocate recommendation resources more purposively, a method named PIM+RA is proposed. This method utilizes a bipartite network that incorporates self-connecting edges and weights. Furthermore, an improved Pearson correlation coefficient is employed for better redistribution. The evaluation of PIM+RA demonstrates a significant enhancement not only in accuracy but also in coverage, diversity, and novelty of the recommendation. It leads to a better balance in recommendation frequency by providing effective exposure to long-tail items, while allowing customized parameters to adjust the recommendation list bias.

## 1 Introduction

As Internet information technology continues to advance rapidly, people have diversified their preferences in obtaining information. However, the accumulation of massive data has also created the problem of information overload (Hemp 2009; Roetzel 2019): on the one hand, users cannot obtain valuable information in a short time; on the other hand, the network content and information produced by information producers are also difficult to accurately present to users who like this content. In addition, information overload can also make users feel fatigued and anxious when shopping online (Ding et al. 2017).A personalized recommendation mechanism is considered as an effective way to solve the problem of information overload (Aljukhadar et al. 2012; Borchers et al. 1998; Fayyaz et al. 2020; Nguyen et al. 2018), which can help users discover the items they are interested in. It has been applied to the vast majority of e-commerce platforms (Alamdari et al. 2020).

Compared with information filters such as search engines, recommendation systems do not require explicit user demand information but instead adaptively establish interest models based on users' historical behavior data. They actively provide users with customized recommendations based on the model results and the users' interests, delivering results and services tailored to their profiles (Feng et al. 2020; Jannach et al. 2021; Shokeen & Rana, 2020). Therefore, the personalized recommendation not only meets users' own information needs but also satisfies the sales needs of merchants, allowing them to dig out users' hidden consumption intentions and better sell products or services, ultimately bringing huge business profits to enterprises(Jannach and Jugovac 2019). In recent years, in addition to the e-commerce field, recommendation systems have been widely used in recommending movies (Thakker et al. 2021), music (Lozano Murciego et al. 2021), books (Ekstrand et al. 2018), news (Raza and Ding 2021), travel routes (Hamid et al. 2021), and other fields.

The recommendation algorithm is the core element that affects the effectiveness of personalized recommendation systems. The vast majority of research work revolves around optimizing and improving. Currently, the mainstream recommendation algorithms mainly include: (1) content-based recommendation (Lops et al. 2019); (2) collaborative filtering recommendation algorithm (Su & Khoshgoftaar, 2009); (3) network-based recommendation (Yu et al. 2016; Zhou et al. 2008); (4) model-based recommendation (Wang et al. 2021); (5) hybrid recommendation (Chang and Jung 2017). Each recommendation algorithm has a different applicable scope. However, most of the current recommendation algorithms still select popular items based on users' preferences after obtaining their portraits, which demonstrates to be inefficient. The Pareto principle (Ge et al. 2022; Lin et al. 2019; Xie et al. 2021) states that popular items stems from their plentiful recommendation resources (Boratto et al. 2021);As non-popular items received little user engagement through purchase or rating, recommendation systems face challenges in recommending them based on the limited interaction information, like browsing and clicking, thereby leading to poor visibility among users (Sreepada and Patra 2021), Existing research has shown that the benefits of recommending long-tail items may exceed those of popular items (Chen et al. 2021; Zhao & Pi, 2019). The benefits may include gaining higher customer loyalty in niche areas, higher profit margins, and filling market gaps. A typical example is Amazon, whose sales data shows that the sales of popular books have been decreasing year by year, while the sales proportion of long-tail books has been increasing(Brynjolfsson et al. 2010; Øverby and Audestad 2021). Amazon has sensitively captured the importance of long-tail recommendations and become one of the world's largest online retailers by constantly iterating its recommendation technology. It is regarded as a pioneering company in the long-tail strategy[1]. Hence, the recommendation system must address the pressing challenge of enabling users to locate high-quality long-tail items and receive recommendations for the same (Smyth and McClave 2001).

Some recommendation algorithms draw ideas from the field of statistical physics. To allocate recommended resources to long-tail items, they apply the dynamic process of resource diffusion to the allocation of recommendation resources. Most resource allocation methods operate in bipartite networks and usually only modify one step in the resource diffusion process (Mumin et al. 2022; Wang and Han 2020; Yin et al. 2020; Zhang et al. 2020). However, optimizing the entire resource allocation process may further enhance the performance of recommendations. At the same time, two main challenges arise with resource allocation methods. First, these methods typically rely exclusively on user and item rating data, highlighting a critical need to optimize their effectiveness. Second, ensuring a universally optimal performance across diverse recommendation scenarios presents a challenge when proposing a specific resource allocation method.

To address these two issues, this paper proposes a network resource allocation recommendation method called PIM+RA. Firstly, the conventional structure of the bipartite network is scrutinized and augmented by interconnecting objects on both the user and item nodes. This approach increases the amount of information transmitted through the bipartite network. It is feasible to further incorporate the item similarity matrix into the model to make resource allocation more purposeful and targeted. Secondly, this work analyzes the deficiencies of the cosine and Pearson correlation coefficient similarity calculation measures and improves the latter based on an analysis of the real dataset scenario. It guarantees the effectiveness of the recommendation information transmitted on the bipartite network. Finally, a resource walking strategy that operates on a bipartite network is proposed, which integrates user and item degree with rating information to precisely aim the final recommendation towards users' preferred items. The experimental results indicate that this strategy can achieve universal performance leadership in different recommendation scenarios.

To validate the effectiveness of the PIM+RA algorithm, multiple evaluations of PIM and PIM+RA are conducted on three real and diverse datasets. The evaluation metric is not limited to accuracy but also includes coverage, diversity, and novelty. The experimental results show that the calculation results of user and item similarity in PIM are more accurate than those of the two traditional algorithms. The PIM+RA algorithm, which incorporates PIM, performs better than the baseline algorithm on multiple indicators under different recommendation list lengths, effectively achieving accurate, diverse, and novel recommendations. In particular, the preference of the recommendation list can be customized by adjusting the parameter, and the change in parameter dependence is discussed in detail in the experiment. Our work provides ideas on how to design resource allocation to improve the comprehensive performance of personalized recommendations in scenarios where only generic rating information is available.

The following contents of this paper are organized as follows: Section 2 overviews the related work of network-based recommendation algorithms, similarity measurement, and long-tail problems. Section 3 introduces the Pearson correlation coefficient similarity improvement, bipartite network improvement, and PIM+RA algorithm in detail. Section 4 introduces the datasets, processing methods, and experiment-cutting process used in this paper. Section 5 introduces the main results of the experiment. Section 6 concludes and discusses the entire paper.

## 2 Related work

### 2.1 Recommendation algorithm based on bipartite network

When adopting object-oriented thinking, the redundant attribute information of users and items can be ignored and abstracted into two nodes, respectively. These nodes are then connected by the interaction data between them, forming a bipartite network structure (Wu et al. 2021). A bipartite network is a type of graph in data structures, where the interaction matrix of $m$ users and $n$ items is abstracted into a $m \times n$ network structure, and edges are connected according to the interaction records to form a bipartite network：

$$BN = \left\langle U^m, I^n, E \right\rangle, \tag{1}$$

where $U^m$ and $I^n$ respectively represent the user and item sets, specifically represented as $U^m = \left\{ u^1, u^2, u^3, ..., u^m \right\}$ and $I^n = \left\{ i^1, i^2, i^3, ..., i^n \right\}$. E is the edge set, which can be represented as $E = \left\{ e_{i^1}^{u^1}, e_{i^2}^{u^1}, e_{i^1}^{u^2}, e_{i^2}^{u^2}, ..., e_{i^n}^{u^m} \right\}$. Many algorithms in graph theory can be adapted for use in recommendation systems, forming the basis of network-based recommendation algorithms. Taking the classic mass diffusion algorithm (MD) as an example, which is based on the principle of material propagation in physics, was first proposed by Zhou et al. (2007). The algorithm considers the user's preference for items as the initial resource. After being multiple equal assigned through the edges, the amount of resources received by the item node is used as the basis for the recommendation. The primary method to generate a recommendation list utilizing the bipartite network involves the following steps:

Step 1: initialize the allocation of resources. the interacted item $i$ for the target user $u$ is allocated 1 unit of initial resources to, and non-interacted items have no initial resources, that is:

$$res_i^u = e_i^u = \begin{cases} 1, & \text{when there exists an edge between user } u \text{ and item } i \\ 0, & \text{when there is no edge between user } u \text{ and item } i \end{cases}, \tag{2}$$

where $res_i^u$ is the initial resource allocated by the user $u$ to the item $i$.

Step 2: resource backtracking. Resources is backtracked from items to users. The resources owned by the items can be evenly distributed to users who interact with them. The resources obtained by the user  are the total resources allocated to all interacting items.

$$res^v = \sum_i \frac{res_i^u \cdot e_i^v}{|U_i|}, \tag{3}$$

where $U_i$ represents the set of users who are connected to the item $i$.

Step 3: resource diffusion. Resources are diffused from users to items. Similarly, a user's resources can be evenly distributed to interacting items. The resources obtained by the item are the total resources allocated to it by all users.

$$res_j = \sum_v \frac{e_j^v \cdot res_v}{|I_v|},\qquad(4)$$

in the above equation, $I_v$ represents the set of items connected to the user $u$.

Step 4: generate a recommendation list. The list is arranged in descending order, according to the resources obtained by each item in the third step. Items used by the user are removed.

In the process of resource allocation, the system upholds the law of conservation of energy, thereby neither increasing nor decreasing the degree of items. Instead, it simply redistributes the available resources. The MD algorithm partly alleviates the problem of data sparsity and has considerable recommendation accuracy. However, due to the inherent advantage of highly popular items (i.e., popular items) in resource allocation, they are more likely to obtain energy, and thus the long tail problem in recommendations has not been addressed.

The subsequently proposed heat conduction algorithm (HC) adjusted the resource allocation method, greatly increasing the exposure proportion of long-tail items (Zhang et al. 2007). However, its recommendation accuracy performance was poor, making it nearly impractical for commercial applications. To balance the performance of recommendations, many scholars have proposed new resource diffusion strategies. For example, Zhou et al. (2008) proposed heterogeneity in initial resource allocation, which makes the initial allocation of resources no longer evenly distributed. Liu et al. (2009) and Lü et al. (2011) proposed two methods of introducing edge weights in resource allocation, in this way, bias will be generated in the walking process of recommended resources in the network. Zhou et al. (2010) proposed a hybrid method combining MD and HC, to balance the accuracy and novelty of bipartite networks. Nie et al.(2015) improved the combination process of Zhou et al. (2010) and proposed the BD algorithm which improve the recommendation comprehensive performance further.

In previous research, many attempts have been made to make local improvements and combinations based on the initial resource allocation methods. However, it has always been challenging to balance the accuracy and innovativeness of recommendations.

## 2.2 Similarity measures in recommendation

Similarity is the result of quantitatively measuring how similar users or items are in a system and is one of the core steps used in link prediction, clustering, and recommendation. Thus, researchers have proposed many similarity measurement methods, such as cosine similarity, Pearson correlation coefficient similarity, Manhattan distance similarity, Euclidean distance similarity, and Jaccard similarity coefficient (Bag et al. 2019; Feng et al. 2019; Gao et al. 2017; Sarwar et al. 2001; Uyanik & Orman, 2023).

Similarity is widely used in collaborative filtering algorithms for recommendation. The core idea of collaborative filtering is to evaluate the similarity between users or items by obtaining user behavior on items, after finding similar neighbors, recommendations are made based on their similarity (Sreepada & Patra, 2021; Su & Khoshgoftaar, 2009). To predict the rating of an unknown item $i$ for the user $u$, we can calculate the similarity matrix and specify the number of nearest neighbors, then utilize the Eq. (5) to make predictions.

$$s_{ui} = \frac{\sum\limits_{v \in kNN(u), s_{vi} \neq \varnothing} sim(u,v) s_{vi}}{\sum\limits_{v \in kNN(u)} sim(u,v)}, \tag{5}$$

where $s_{ui}$ represents the rating of the user $u$ for the item $i$, $kNN(u)$ represents the set of the nearest neighbors of user $u$, and the optimal number of neighbors can be selected through model training; $sim(u,v)$ represents the similarity between the users $u$ and $v$.

Therefore, similarity measures will greatly affect the effectiveness of recommendations, not only in collaborative filtering but also in other recommendation scenarios that require similarity (Fkih 2022). Representative work involves the selection of appropriate similarity measurement methods under cold-start conditions, which is critical to improving the recommendation accuracy of the electronic learning recommendation system (Joy and V.G. 2020).

Scholars have improved the similarity measures in recommendation systems to suit various recommendation scenarios. To better capture user behavior patterns and preferences, hybrid similarity has been enhanced by incorporating implicit feedback and this has led to an improvement in the quality of recommendation systems (Liu et al. 2017). Vectorized similarity can measure the similarity between users across multiple dimensions based on the characteristics of the item (Su et al. 2020). Considering user social trust networks and location-aware similarity measurements can improve recommendation prediction accuracy in specific scenarios (Gao et al. 2017; Zeng et al. 2022). Combining resource allocation index with cosine similarity can improve the accuracy, diversity, and novelty of recommendation systems (Chen et al. 2017), and so on.

Many similarity improvement measures require the integration of additional information. However, there is limited research on how to effectively improve similarity measure when information is lacking. Furthermore, few scholars have proposed corresponding strategies for using improved similarity in resource allocation.

**2.3 Long-tail problems in recommendation**

Currently, most recommendation systems select popular items according to the user's preferences after obtaining the user's portrait through their characteristics, and such recommendations are often inefficient because users are likely to already be familiar with popular items. The economic law of the "Pareto principle" may explain this phenomenon: a large number of ratings and interactions in similar products in the market are only included in a few items, which are classified as popular items. Due to the large number of ratings and interactive information that popular items have, traditional recommendation methods tend to recommend popular items while ignoring non-popular ones due to algorithm mechanisms (Boratto et al. 2021; Steck 2011). These non-popular items thus enter a vicious cycle where users do not purchase, rate, or interact with them, leading to a lack of data to base recommendations on and ultimately causing users to be unable to see them (Sreepada and Patra 2021). This is the long-tail problem in recommendation systems.

The long-tail problem is an economic phenomenon that stems from numerous statistical results and affects personalized recommendation systems within the realm of e-commerce. Addressing this issue is crucial in facilitating consumer access to high-quality long-tail products and generating relevant recommendations for such items (Zhao and Pi 2019). Recent studies have also noticed the long-tail issue in recommendations. Chen et al. (2021) designed the HashCF tourism recommendation system, which combines power law distribution and locality-sensitive hash to solve the low diversity and severe data sparsity problems in tourism recommendations. Huang and Wu (2019) proposed a framework consisting of project summary extractor and similar project extractor components. By identifying online reviews and consumer emotions, an enhanced k-medoids method is used to identify equivalent long-tail substitutes and recommend them to users for purchase decisions. So far, effective methods for long-tail recommendations are still being explored.

## 3 Method design

Existing literatures show that the MD algorithm improves the ranking of high-popularity items in the recommendation list. However, it performs poorly in terms of coverage and novelty (Nie et al. 2015). Item-based collaborative filtering recommendation (IBCF) focuses on recommending items from the similar neighborhood of historical items interacted by users, thus achieving strong long-tail item recommendation capability, but it performs poorly in terms of accuracy (Thakkar et al. 2019). The bipartite network stores the node information of users and items, which can record the similarity of items in IBCF, enabling collaborative filtering. Combining these two algorithms is expected to achieve a balance between accuracy and other metrics. Since the recommendation effectiveness of collaborative filtering algorithms depends heavily on the similarity measure, in the following discussion, we will design an improved similarity measure will be designed and applied it to resource allocation in the bipartite network. The framework of this method is shown in **Fig. 1**.

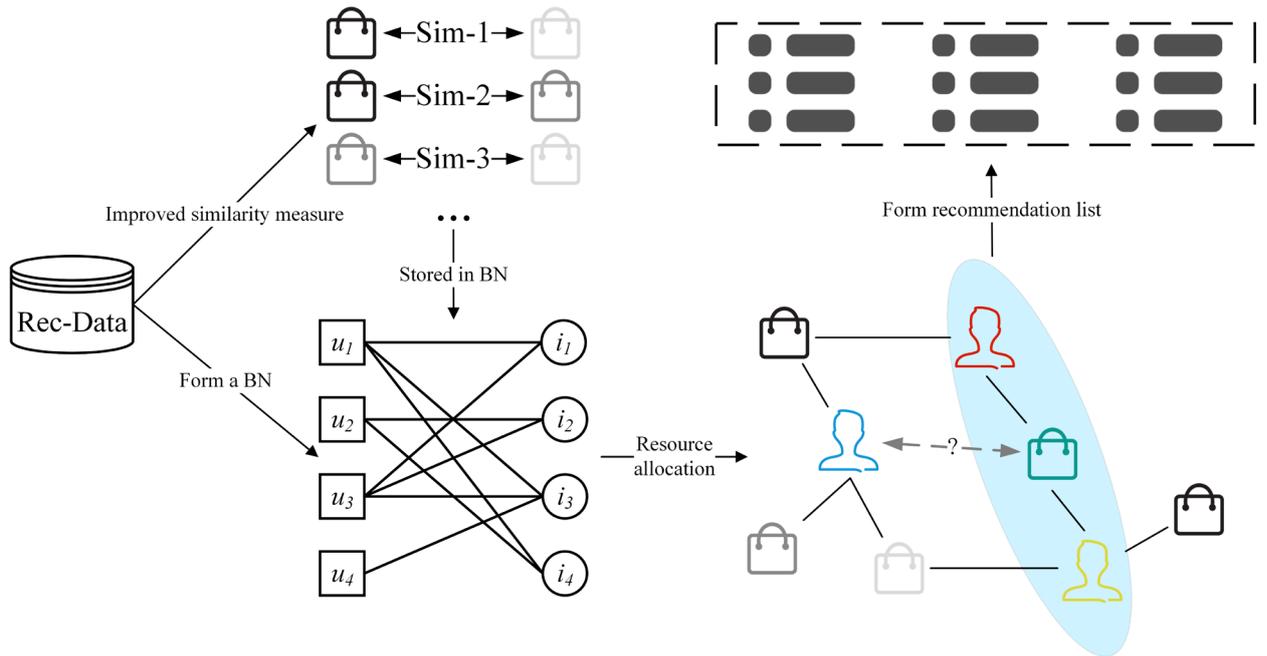

**Fig. 1.** Method framework

### 3.1 Improvement of Pearson correlation coefficient (PIM)

The traditional Pearson correlation coefficient (PCC) based similarity measure is briefly introduced first. Due to users' rating styles, some prefer to rate low while others prefer to rate high. Simply using their co-rating-item (CRI) for similarity calculation often leads to significant estimation errors (Wang & Fu, 2021). The PCC-based similarity measure can better avoid this problem, and is defined as follows:

$$sim_{pcc}(u,v) = \frac{\sum\limits_{i \in I_u \cap I_v} \left( s_{ui} - \overline{s_u} \right)\left( s_{vi} - \overline{s_v} \right)}{\sqrt{\sum\limits_{i \in I_u \cap I_v} \left( s_{ui} - \overline{s_u} \right)^2 \sum\limits_{i \in I_u \cap I_v} \left( s_{vi} - \overline{s_v} \right)^2}}, \tag{6}$$

where $s_{ui}$ represents the rating of the user $u$ for the item $i$, $\overline{s_u}$ represents the average rating of the user $u$ for all items, $I_u$ and $I_v$ represent the sets of items rated by users $u$ and $v$, respectively. $sim_{pcc}(u,v) \in [-1,1]$.

In collaborative filtering algorithms, the higher the computed similarity, the more likely the two users are to share similar interests. However, considering the increasing problem of matrix sparsity, the original similarity measure may have errors in measuring the true similarity between the two due to the small number of CRI. Taking the scoring matrix on the left side of **Fig. 2** as an example, the similarity of three neighbors of the user $u_3$ is calculated using cosine (CS) and PCC based similarity measure, respectively, and the results are obtained as shown in the table on the right side:

| | $i_1$ | $i_2$ | $i_3$ | $i_4$ |
|---|---|---|---|---|
| $u_1$ | 5 | - | 3 | 2 |
| $u_2$ | - | 4 | 3 | - |
| $u_3$ | 4 | - | 1 | 2 |
| $u_4$ | - | 3 | - | 5 |

| Neighbor | CS | | PCC | |
|---|---|---|---|---|
| | Similarity | Ranking | Similarity | Ranking |
| $u_1$ | 0.956 | 1 | 0.786 | 2 |
| $u_2$ | 0.131 | 3 | 1.000 | 1 |
| $u_4$ | 0.374 | 2 | -1.000 | 3 |

**Fig. 2.** Comparison of CS and PCC calculation methods

As can be seen from **Fig. 2**, for the same user, the two similarity calculation methods yield almost completely different results. The calculation bias of the algorithm is caused not only by the different measures but also by the extreme sparsity of the scoring matrix and the problem of too few user intersections (Wang et al. 2022). To solve this problem, an improved similarity measure PIM, which is based on the PCC is proposed. PIM considers user activity, item popularity, and CRI, the aim of these is to enhance the accuracy of collaborative filtering. The following sections will be explained from the perspective of user similarity.

### 3.1.1 CRI rewards and punishments

The traditional PCC measure accounts only for the CRI of users, but the user rating matrix in most platforms is excessively sparse, resulting in a small number of CRI. This sparsity negatively affects the accuracy of the calculation, leading to an inadequate user neighborhood selection. In consequence, the obtained user neighborhood may not necessarily reflect the real neighborhood of users. The same is true for the calculation of the set of neighborhoods of items. **Fig. 3** illustrates the distribution of CRI ratios in the three datasets.

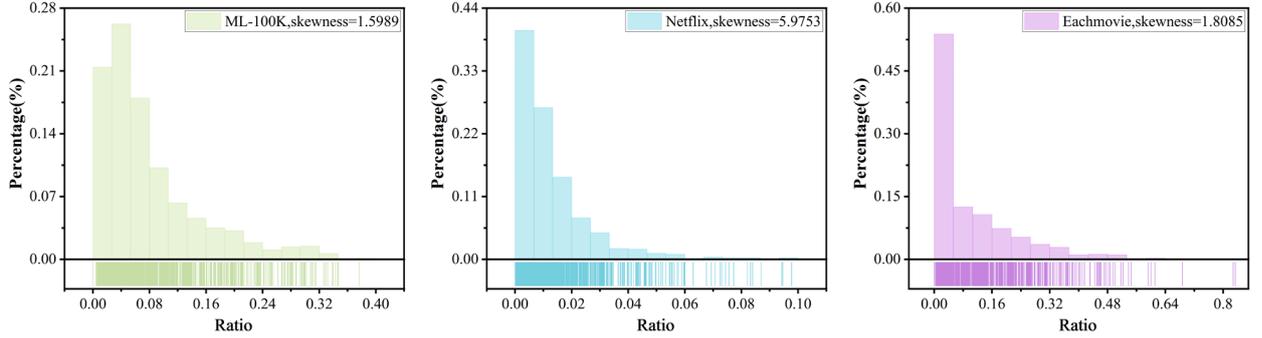

**Fig. 3.** Ratio of intersection to union distribution of user CRI. 50 users are randomly selected from the datasets and the ratio of intersection to union of any two users is counted.

As shown in **Fig. 3**, in the three datasets, the ratio of intersection to union of users' CRI shows a significant right-skewed trend, i.e., skewness > 0.

If the count of two users' CRIs is notably higher than the average count in the dataset, their similarity remains high even when they differ in rating specific items. This is because there is a latent layer of user-initiated item selection relationships before rating specific items. When there are only a few CRIs, occasional similar ratings can impact the PCC calculation, and therefore, the former should be promoted, and the latter should be penalized. Define the first improved PCC measure as:

$$sim_{pim}^{1}(u,v) = \ln(1 + \frac{\left| I_u \cap I_v \right|}{\left| I_u \cup I_v \right| \cdot ar}) \cdot sim_{pcc}(u,v), \tag{7}$$

where $ar$ is mean CRI ratio of intersection to union for all user or item groups in a particular dataset. It can be written as:

$$ar = \begin{cases} 2\dfrac{\displaystyle\sum_{u,v\in U, u\neq v}\dfrac{\left|I_u \cap I_v\right|}{\left|I_u \cup I_v\right|}}{\left|U\right|\left(\left|U\right|-1\right)}, \text{when calculating user groups} \\[20pt] 2\dfrac{\displaystyle\sum_{i,j\in I, i\neq j}\dfrac{\left|U_i \cap U_j\right|}{\left|U_i \cup U_j\right|}}{\left|I\right|\left(\left|I\right|-1\right)}, \text{when calculating item groups} \end{cases}, \tag{8}$$

### 3.1.2 Active user penalty

Active users interact with a higher number of items and are more likely to have item intersections with other users than common users. In an extreme case, the number of CRIs for an active user and an inactive user is exactly equal to the number of rating items for the inactive user. As shown in **Fig. 4**, the average popularity of interacted items is higher for inactive users, and they are likely to be in the active user's item interaction list, but it does not necessarily mean that the two users are more similar.

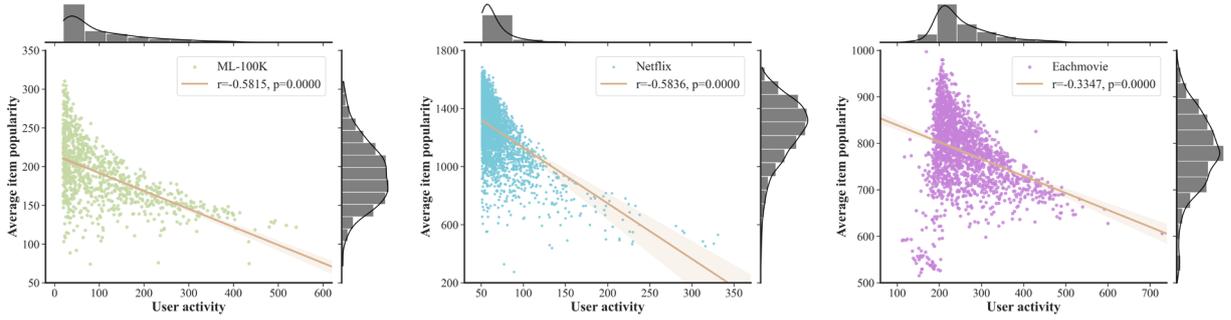

**Fig. 4.** Relationship between user activity and average item popularity. Each dot in the graph represents the average popularity of items interacted with by the user at that user activity level. Regressions are fitted to all points to obtain the regression coefficients, p-values, and regression lines.

In addition, users are rewarded and punished based on the number of CRIs in Section 3.1.1. However, the punishment hardly worked for active users. The reason for this is as described above: the list of rating items is far more inclusive for active users than for common users. Therefore, the improved PCC for the active user penalty is defined as follow:

$$sim_{pim}^2(u,v) = \frac{sim_{pim}^1(u,v)}{1+e^{1-\frac{\max\limits_{u'\in U}\{|I_{u'}|\}}{|I_u|+|I_v|}}}, \tag{9}$$

The definition takes into account the activity of the two users in the user pair. The similarity between two individuals is weakened correspondingly when one is extremely active or both are relatively active.

### 3.1.3 Popular item weighting correction

When a movie is very popular, almost all users will watch it; when a book is regarded as a classic, almost all users will read it. The popularity of an item is often related to its quality, as shown in **Fig. 5**, the more popular the item, the higher its rating tends to be. The PCC-based similarity measure uses the average user score as a benchmark for the preference measure, but since most users rate popular items highly, such items contribute less to the similarity calculation. The traditional similarity measure treats all items equally, i.e., assigns equal weight to them. To reduce the influence of popular items on the similarity calculation, a weight correction factor can be added to the similarity calculation to balance the influence of popular and non-popular items on the similarity calculation.

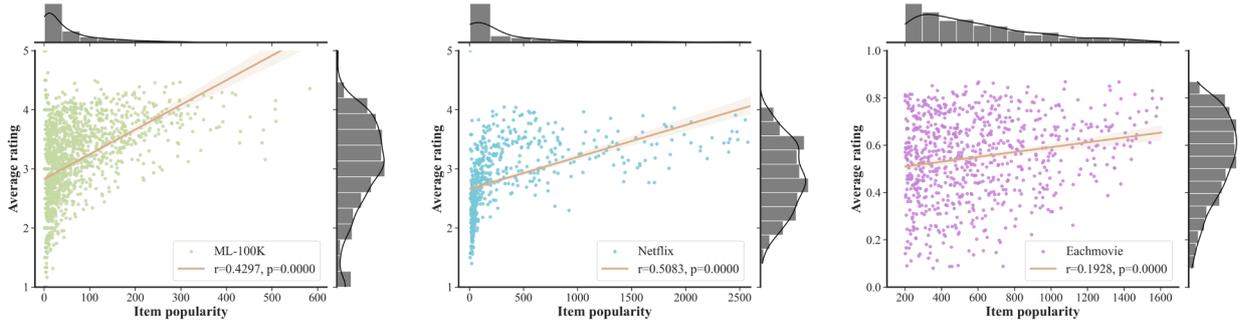

**Fig. 5**. Relationship between item prevalence and mean rating. Each dot in the graph represents the corresponding mean rating at that item's popularity level. Regressions were fitted to all dots to obtain regression coefficients, p-values, and regression lines.

Combining the improved definitions of PCC-based similarity measures in Eqs.（7）and（9），the final PIM measure is defined as follows:

$$sim_{pim}(u,v) = \frac{\ln(1 + \frac{|I_u \cap I_v|}{|I_u \cup I_v| \cdot ar}) \cdot \sum\limits_{i \in I_u \cap I_v} \frac{1}{\lg(1 + |U_i|)} \left(s_{ui} - \overline{s_u}\right)\left(s_{vi} - \overline{s_v}\right)}{\sqrt{\sum\limits_{i \in I_u \cap I_v} \left(s_{ui} - \overline{s_u}\right)^2 \sum\limits_{i \in I_u \cap I_v} \left(s_{vi} - \overline{s_v}\right)^2} \cdot (1 + e^{1 - \frac{\max\{|I_u|\}}{|I_u| + |I_v|}})}.$$

(10)

### 3.2 Resource allocation method with PIM integration

At first, resource allocation algorithms applied to bipartite networks often do not consider edge weights, which results in a uniform allocation of resources to objects with adjacent vertices. However, the rating information can reflect how much users like the items and can be considered one of the bases for resource allocation. Meanwhile, to make the resource random walking process on the bipartite network addable to the pre-node information for similarity transfer, the original bipartite network structure is optimized by weighting the original connected edges with user ratings, and by adding self-connecting edges to nodes. The optimization is shown in **Fig. 6**

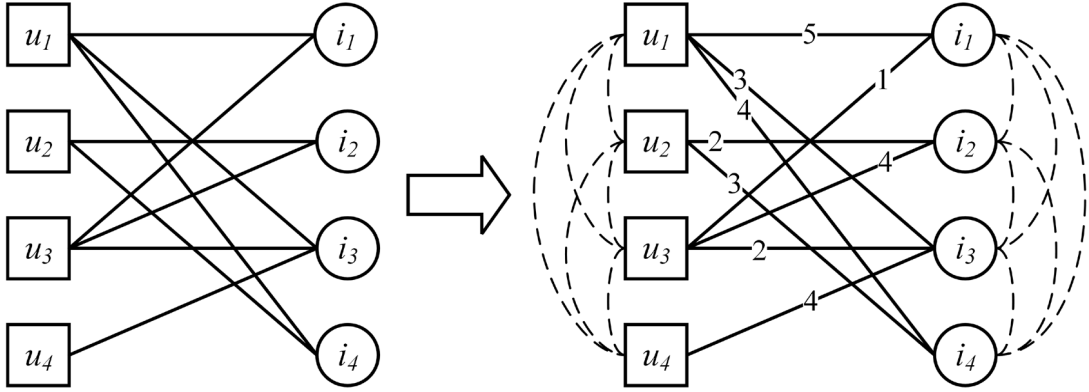

**Fig. 6.** Improvement of bipartite network. The user-item interaction information in this figure is consistent with the rating matrix in **Fig. 1**.

Regarding the improved bipartite network, its structure includes the following:

$$BN^{IM} = \left\langle U^m, I^n, E_w, E_u, E_i \right\rangle \tag{11}$$

where the meaning of $U^m$ and $I^n$ the same as Eq. (1), respectively represent the user and item sets. $E_w$, $E_u$ and $E_i$ respectively represent weighted edges, user self-connecting edges, and item self-connecting edges. The above elements constitute an improved bipartite network.

According to previous studies, it appears that because popular items are frequently interacted with by a significant number of people, they have the potential to recommend fewer users than long-tail items. The recommendation of long-tail items is directly related to the diversity, novelty, and coverage of recommendation results, and small-degree items tend to be more recommendable than large-degree items (Abdollahpouri et al. 2019). Therefore, the three-step resource random walking process will be modified on the original bipartite network by commencing with item degree and user degree, to reveal mechanisms that enable small-degree items to obtain accurate recommendations.

Inspired by algorithms such as MD, HC, and PageRank (Park et al. 2019), the processes of resource initialization, resource wandering, and resource acceptance are modified in the resource allocation process to achieve a balance of accuracy and diversity.

For the resource initialization process, highly active users should be allocated more initial resources when they choose small-degree items. Since small-degree items mostly interact with active users (see **Fig. 3**), the allocation of initial resources will be more effectively diffused based on the selection relationships of active users; when low-active users select large-degree items, it is often interpreted as emulating behavior, which may not accurately reflect their true preferences, therefore, the initial resource allocation should be reduced. The resource initialization process is

$$R^1_{(u \to i)} = \frac{1}{|I_u|} + \ln\left(\frac{|I_u|}{|U_i|}\right), \tag{12}$$

To avoid significant differences in initial resources between highly active and less active users, the factor $1/|I_u|$ is included as a supplement to the initial resources as a user activity factor. Subsequently, the resources acquired by the items are allocated to the users proportionally according to the weights on the edges:

$$R^2_{(i \to v)} = \frac{R^1_{(u \to i)} \cdot w_{vi}}{w_i}, \tag{13}$$

where $w_{vi}$ denotes the rating of the item $i$ by the user $v$, and $w_i$ denotes the sum of ratings of all connected edges of the item $i$. In the final resource-receiving session, considering the strong resource acquisition ability of the large-degree items, even after balancing the initial resource allocation, more resources will still flow into the large-degree items from the connected edge. The resource accessibility of items is therefore revised to balance popular items with long-tail items. At the same time, resources will be allocated in a more targeted manner based on the similarity of the pre-node item to the back-node item:

$$R^3_{(v \to j)} = \frac{R^2_{(i \to v)} \cdot w_{vi} \cdot sim_{pim}(i, j)}{|U_j|^{\theta} \cdot w_{v.}}, \tag{14}$$

where $\theta$ is an adjustable parameter with the interval $[0,1]$. When $\theta$ is set larger, the weakening of resource accessibility to large-degree items is greater; when $\theta = 0$, it means no correction is made to it. The size of $\theta$ can be defined according to the actual demand for the diversity of the recommendation list. Meanwhile, the resource allocation in Eq. (14) employs the PIM method to participate in collaborative filtering which enhances stability while broadening recommendation outcomes.

It is worth mentioning that, despite the addition of self-connecting edges for both users and items, in our tests, bipartite network incorporating IBCF achieves more stable and diverse recommendation results than incorporating user-based collaborative filtering recommendation (UBCF). While incorporating UBCF into bipartite network greatly extends the novelty of the recommendation list, it also leads to a significant decrease in accuracy. The reason for this phenomenon is due to the interpretability of IBCF and the ease of accurately measuring item similarity. Therefore, this paper only considers incorporating item similarity in the resource random walking process. The specific algorithm flow schematic is shown in **Fig. 7**

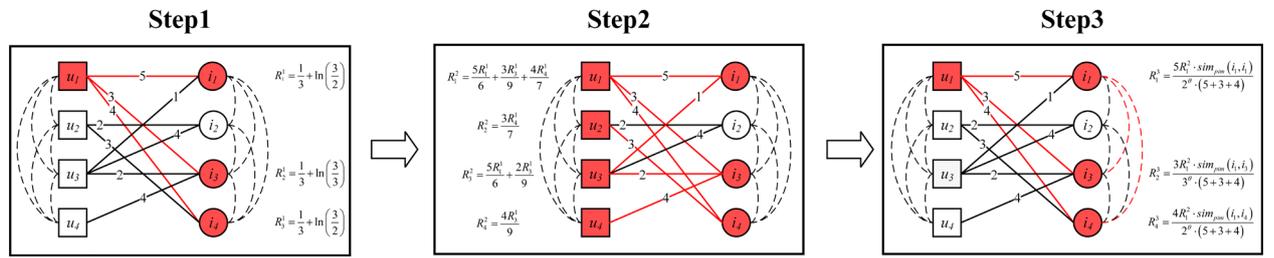

**Fig. 7.** A schematic representation of the PIM+RA algorithm flow

*Note*: The scoring matrix in the figure is the same as in **Fig. 1**. The process takes the user $u_1$ as the recommendation object, and Step 1 initializes the resources for the three items it has interacted with; Step 2 backtracks the resources for the users who have interacted with the three items in turn and calculates the total resources obtained by each user; Step 3 takes the user $u_1$ as an example and assigns the resources obtained in Step 2 to the items with which it has interaction according to the assignment rules. By traversing the process of Step 3 for all users and sorting resources by item, the recommendation list for the user $u_1$ can be obtained.

## 4 Experimental setting

### 4.1 Data and processing

Three datasets with unique characteristics are selected to evaluate the performance of recommendation algorithms. The relevant descriptions of the three datasets are as follows:

(1) ML-100K

Movielens is a non-commercial, personalization-based movie recommendation website and also an experimental site for research purposes, containing a huge amount of basic information about movies and users as well as corresponding rating data[1]. The 100K rating data captured by

---

1. https://movielens.org/

GroupLens Research on this website (Harper and Konstan 2015) was used, and the dataset will be referred to as ML-100K in this paper.

(2) Netflix

The Netflix dataset was built to support participants in the 2006 Netflix Recommendation Contest, owned by Netflix, the world's largest online movie rental service platform[1]. The raw dataset covers 480,000 users and 17,000 movies, totaling about 1 million quantitative ratings.

(3) Eachmovie

The DEC Systems Research Center ran the Eachmovie recommendation service for 18 months to experiment with collaborative filtering algorithms[2]. During this period, they collected a total of 72,916 users and 1,628 movies, with a combined total of about 2.8 million ratings.

To simulate the formation of recommendation environments with different densities and user-item ratios, the datasets are processed. Specifically, any processing is not done on ML-100K; for Netflix, the top 600 movies by number are extracted and user ids with more than 50 rating records are selected as the experimental dataset; for Eachmovie, only the users and items that have more than 200 rating records are filtered out to simulate a denser recommendation environment. **Table 1** describes the basic information from the above three datasets.

**Table 1**

Basic information about the dataset. The three numbers in the scoring interval mean, in order, the lower limit of rating, the upper limit of rating, and the step unit. A higher sparsity indicates a sparser dataset.

| Dataset | Users | Items | Links | Rating interval | Sparsity |
|---------|-------|-------|-------|-----------------|----------|
| ML-100K | 943 | 1682 | 100000 | [1,5,1] | 93.70% |
| Netflix | 3071 | 600 | 213116 | [1,5,1] | 88.43% |
| Eachmovie | 1781 | 771 | 457942 | [0,1,0.2] | 65.49% |

To verify the effectiveness of PIM+RA and avoid the contingency of experimental results, K-fold cross validation (KCV) is used to randomly divide each dataset into K subsets of equal number and mutual exclusion, and K is set to 5.

## 4.2 Metrics

In previous literature, evaluation metrics for recommendation performance have been extensively studied (Avazpour et al. 2014; Fayyaz et al. 2020; McNee et al. 2006). In this paper, five widely used metrics are used to quantitatively evaluate the performance of each recommendation algorithm, including accuracy, coverage, internal diversity, intra-user diversity, and novelty. The following briefly describes these metrics.

---

1. https://movielens.org/
2. https://grouplens.org/datasets/eachmovie/

Ensuring the accuracy of recommendations is the most basic and important task of the recommendation algorithm, and only based on the recommendation accuracy reaching the standard, it is meaningful to pursue other optimization methods of the algorithm. In the top-k recommendation task, ranking accuracy is often used (Fayyaz et al. 2020; Tang et al. 2023). In this paper, average ranking score (ARS) is used as a metric to evaluate ranking accuracy (Krichene and Rendle 2022), and the formula can be defined as

$$ARS = \frac{|U|}{\sum_{u \in U} \sum_{i \in S_f^{Test}} \frac{r_i}{|L_u|}}, \tag{15}$$

where $L_u$ is the recommendation list generated by the recommender system for the user $u$, $S_f^{Test}$ is the set of items that users like in the test set, and $r_i$ is the ranking of the item $i$ in the recommendation list. The ranking score is calculated by separately computing the score for the set of items favored by each user and averaging the scores for all users. As a result, the ARS for the entire recommender system can be obtained. Note that the positions of the numerator and denominator of the original ARS metric are switched, in this paper, the larger the ARS, the higher the accuracy of the recommendation.

Coverage is a metric to characterize the system's ability to recommend long-tail items and the coverage of recommended items (Ge et al. 2010), which can be further divided into prediction coverage, list coverage, and catalog coverage according to different needs (Avazpour et al. 2014). However, they can only measure whether an item has appeared in the recommendation list as a criterion and do not respond well to the number of recommendations for each item, i.e., items that have been recommended multiple times in the recommendation list have the same coverage contribution as items that have been recommended only once. Therefore, the Gini coefficient (Ren et al. 2014) is used as the evaluation criterion for coverage, which is calculated as

$$Gini = 1 - \frac{1}{|I| - 1} \sum_{i \in I} \left( 2r_i - |I| - 1 \right) P(i), \tag{16}$$

where $P(i)$ is the total number of recommendations for the item $i$ divided by the number of recommendations for all items. A higher Gini coefficient indicates broader coverage of recommendations and greater balance of recommendations for each item.

The diversity of recommendations can be further divided into internal diversity (ID), which refers to the diversity of recommended items in the recommendation list generated for one user, and inter-user diversity (IUD), which focuses on the differences between the recommendation lists of different users (Aytekin and Karakaya 2014). ID needs to be calculated in conjunction with similarity, and the formula is shown as follows:

$$ID = \frac{1}{|U|} \sum_{u \in U} \left( 1 - 2 \frac{\sum_{i,j \in L_u, i \neq j} sim(i,j)}{|L_u|(|L_u| - 1)} \right), \tag{17}$$

The ID of the recommender system is obtained by averaging the IDs of all user pairs. And the IUD can be calculated using Hamming Distance (HD) (P. Yin and Zhang 2020), which is calculated as

$$HD(u,v) = 1 - \frac{|L_u \cap L_v|}{|L|}, \tag{18}$$

where $|L|$ is the length of the recommendation list. Clearly, $HD(u,v) \in [0,1]$, when the recommendation lists of the user $u$ and $v$ are identical, their HD is 0, and 1 when they are completely different. Similar to the ID, the IUD of the entire recommender system can be obtained by averaging HD between all pairs of users.

$$IUD = HD_{rs} = 2 \frac{\sum_{u,v \in U, u \neq v} HD(u,v)}{|U|(|U| - 1)}, \tag{19}$$

Recommending popular items is often inefficient because users are likely to already know them, and the reason for this lies in the lack of novelty of the recommendation. It is commonly believed that novel recommendations need to possess some differentiation, i.e., the recommended items are different from items that the user has interacted with before (Castells et al. 2022). Therefore, a novelty calculation metric based on the average distance is adopted, which can take into account the similarity gap between the recommended items and the user's historical interaction items, namely

$$Novelty = 1 - \frac{1}{|U|} \sum_{u \in U} \frac{\sum_{i \in L_u, j \in I_u} sim(i,j)}{|L_u||I_u|}, \tag{20}$$

## 5 Results

To validate our work, the following main research questions (RQs) are formulated to test the validity of the proposed PIM+RA recommendation method:

RQ1: Effectiveness of PIM: Does the improved PIM measure the similarity between users or items more accurately than traditional PCC and CS?

RQ2: Performance of PIM+RA: Does the resource allocation method considering user activity, item popularity, and improved similarity improve the comprehensive recommendation performance?

RQ3: Relationship between the parameter and PIM+RA: For the unique adjustable parameter $\theta$, how does it affect the performance of PIM+RA?

RQ4: Distribution of recommended times for each item by PIM+RA: How does PIM+RA balance the number of recommendations for long-tailed items and popular items?

RQ5: Correlation between PIM+RA performance and recommendation list length: How will the performance of PIM+RA change under different recommendation list lengths, and what is the performance gap relative to the baseline method?

To address the aforementioned questions, we conduct experimental analysis utilizing the ML-100K, Eachmovie, and Netflix datasets, which were previously mentioned in the previous section. Then, comparative experiments are conducted on the above research questions. Finally, the findings from the experiments are discussed in detail.

### 5.1 Effectiveness of PIM (RQ1)

We compare the PIM with the traditional PCC and CS in the UBCF and IBCF scenarios, respectively. The specific process is to first use the above three methods to calculate the user similarity and item similarity on each training set, and then normalize them to rank the user nearest neighbors and item nearest neighbors in descending order of similarity. Subsequently, the top $k(k \geq 1)$ nearest neighbors are sequentially extracted, and Eq. (5) is used to calculate the user's predicted rating for the items in the test set. Finally, the normalized root mean square error (NRMSE) is used as the accuracy evaluation metric to measure the distance between predicted and actual ratings for each method. **Fig. 8** shows the NRMSE comparison results of the three similarity measures.

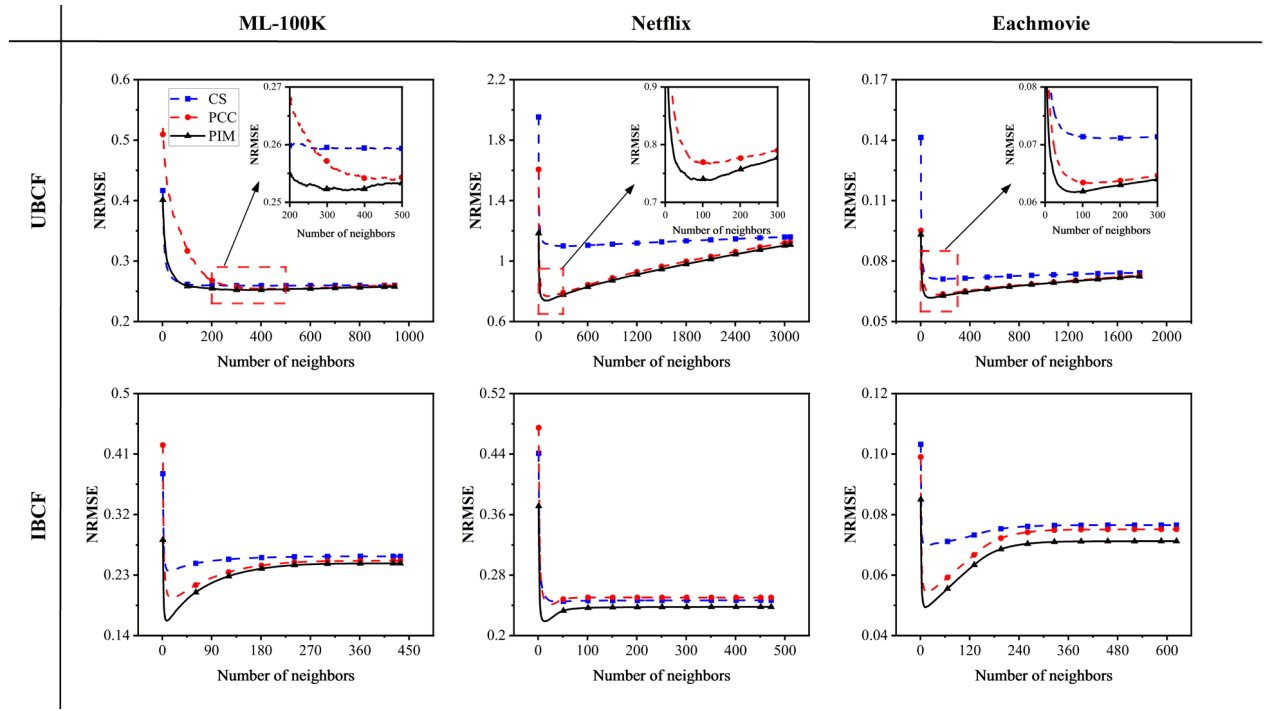

**Fig. 8.** Comparison of similarity measures. The blue, red, and black lines represent the CS, PCC, and PIM measures, respectively. Local enlargement was performed in the UBCF scene.

It can be seen from **Fig. 8** that the PIM measure proposed in this paper can achieve the best accuracy results in all test scenarios, and when the best accuracy rate is obtained, compared with the PCC measure and CS measure with the same k value, both accuracy rates can be improved: in UBCF, the increase is 1.11% to 4.15% and 2.90% to 50.87%; in IBCF, the increase is 10.81% to 27.28% and 15.34% to 47.07%. At the same time, in most scenarios, the accuracy of the PIM measure is better than the other two traditional similarity measurement measures for any same k value.

**Fig. 9** shows the similarity distribution between the PIM measure and the PCC measure. Observing the distribution of both, the similarity measured by the PCC measure is mostly distributed in a range of 0.5 to 1, while the upper and lower densities of the dense bands are unevenly distributed. The PIM measure has significantly improved the problem of the distribution gap, the dense band tends to be closer to the location where the normalized similarity is 0.5, and the upper and lower distribution density are more uniform, showing a normal distribution overall. The above features are more obvious in the ML-100K data set with higher sparsity. A good similarity distribution facilitates the progressive differentiation of users and items while avoiding the impact of extreme values on collaborative filtering, ultimately improving the accuracy and stability of recommendations (Han 2020).

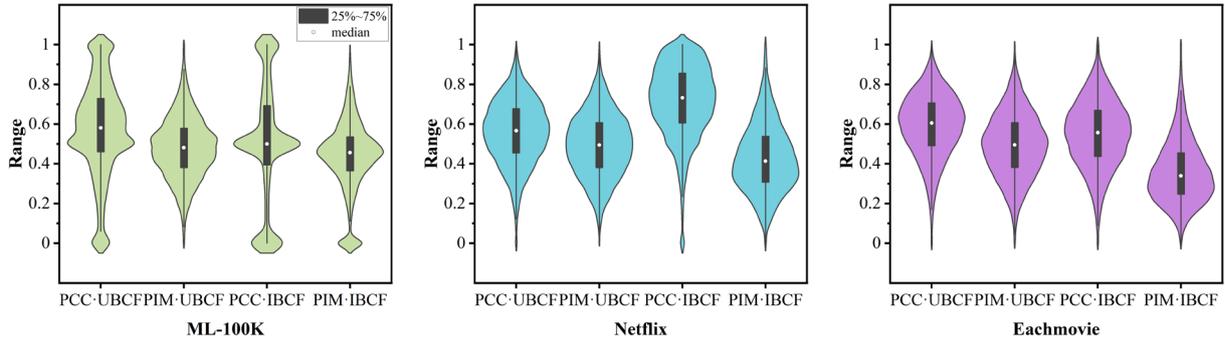

**Fig. 9.** Comparison of PCC and PIM similarity distributions. In the violin plot of each distribution, the white dots represent the median of the distribution, the black rectangles cover the upper and lower quartiles of the distribution, and the width of the graph represents the density of the distribution at that similarity.

## 5.2 Evaluation of PIM+RA

### 5.2.1 Performance (RQ2)

The algorithm PIM+RA was evaluated for all-around performance using three datasets and the five metrics specified in Section 4. Benchmark comparison methods included the UBCF, IBCF, MD, and SVD algorithms. When limiting the length of the recommendation list to 100, the differences in the recommendation effects of several algorithms are compared, and the results are shown in **Table 2**.

**Table 2**

Summary review results for the three datasets. The best-performing entries under each metric are highlighted in bold black. The number after the PIM+RA bracket represents the value of the parameter $\theta$.

| Dataset | Method | ARS | Gini | ID | IUD | Novelty |
|---------|--------|------|------|------|------|---------|
| ML-100K | UBCF | 0.3644 | 0.0725 | **0.5431** | 0.2498 | 0.5029 |
| | IBCF | 0.2382 | **0.4500** | 0.3868 | **0.8710** | 0.4612 |
| | SVD | 0.3356 | 0.1105 | 0.4074 | 0.4844 | 0.4373 |
| | MD | 0.8588 | 0.1143 | 0.4298 | 0.5247 | 0.4307 |
| | PIM+RA (0.6) | **0.9389** | 0.3632 | 0.4964 | 0.8324 | **0.5056** |
| Netflix | UBCF | 0.7793 | 0.1937 | 0.3538 | 0.2262 | 0.4017 |
| | IBCF | 0.3654 | **0.6133** | 0.2477 | **0.7456** | 0.3469 |
| | SVD | 0.6268 | 0.2680 | 0.3182 | 0.4321 | 0.3841 |
| | MD | 1.4641 | 0.2226 | 0.4382 | 0.3264 | 0.4348 |
| | PIM+RA (0.6) | **1.5344** | 0.3591 | **0.5504** | 0.5511 | **0.5928** |
| | UBCF | 0.1066 | 0.1804 | 0.3872 | 0.3595 | 0.4775 |
| | IBCF | 0.1173 | 0.3950 | 0.3803 | 0.6852 | 0.4816 |
| Eachmovie | SVD | 0.1055 | 0.2080 | 0.3926 | 0.4427 | 0.4797 |
| | MD | 0.1610 | 0.2794 | 0.4741 | 0.5864 | 0.4855 |
| | PIM+RA (0.9) | **0.1913** | **0.5341** | **0.5519** | **0.7821** | **0.6065** |

It can be found that the PIM+RA algorithm obtains the best results in 10 of the five metrics in the three datasets, and the overall performance is better than the other algorithms. Among them, the accuracy and novelty of PIM+RA on the three datasets are better than the other four algorithms, which are (0.9389, 0.5056), (1.5344, 0.5928), and (0.1913, 0.6065).

Noting that in the Eachmovie dataset, PIM+RA outperforms the other four algorithms in all metrics; in the ML-100K and Netflix datasets, although PIM+RA cannot guarantee that all metrics outperform the other algorithms, it still maintains a balance of other metrics with high accuracy. For Gini and IUD, PIM+RA lags behind IBCF but outperforms the other three algorithms; for ID, it lags behind the UBCF algorithm by about 9.4% in the ML-100 dataset, but this disadvantage is relatively small compared to the accuracy gap between them (0.3644 and 0.9389, respectively), and it outperforms UBCF in the ID evaluation of the other two datasets. Overall, PIM+RA achieves excellent accuracy and novelty, while having richer diversity and wider coverage. In addition, the bias of the recommendation list can be modified by adjusting the parameter $\theta$, and it will be discussed later.

### 5.2.2 Relationship between the parameter and PIM+RA (RQ3)

The performance metrics and the average item popularity of the recommendation list obtained by the PIM+RA algorithm are closely dependent on the parameter $\theta$. Further, **Fig. 10** plots the relationship between the metrics and $\theta$ for the PIM+RA algorithm in the three datasets. It is observed that with the increase of $\theta$, PIM+RA will tend to recommend long-tail items, and the popularity of items in the recommendation list will gradually decrease, while Gini, IUD, and Novelty will be greatly improved. Among them, Gini and IUD converged when $\theta$ was close to 1, while Novelty increased significantly when $\theta \geq 0.7$. In the three datasets, for each 0.1 increase in $\theta$, the average decrease in item popularity is 17.0%, 9.3%, and 5.9%, respectively, with Gini, IUD, and Novelty improving on average (13.5%, 10.3%, 4.6%), (4.0%, 7.3%, 1.8%), and (1.2%, 0.6%, 0.4%). Based on this, long-tail items received more recommendation opportunities. Within a certain interval, ARS will increase as the percentage of long-tail items in the recommendation list increases and peaks around $\theta = 0.5$. When $\theta = 0.5$, the ARS in the three datasets increased by 4.5%, 2.5%, and 13.0%, respectively, compared to $\theta = 0$.

Based on the above findings, it can be concluded that cold items have stronger recommendation ability, and adding a certain degree of cold items to the recommendation list can enhance the attractiveness to users. However, when the number of cold items gradually increases, it is difficult to guarantee the accuracy of the recommendations, and there is a situation where the stability of the recommendations is given up to highlight the novelty of the recommendations, which is reflected in the decrease in accuracy after $\theta \geq 0.5$. For internal diversity, no uniform pattern emerges across datasets.

It is worth noting that even if the $\theta$ is set to 0 without penalizing the popular item's ability to receive resources, the PIM+RA algorithm still substantially optimizes the performance of the recommendation list on various performance metrics. Taking the Eachmovie dataset as an example, the ARS, ID, and Novelty of PIM+RA (0) are superior to all other algorithms, and its Gini and HD are (0.3270, 0.6471), slightly inferior to IBCF (0.3950, 0.6852), but significantly better than UBCF, SVD, and MD which are (0.1804, 0.3595), (0.2080, 0.4427), and (0.2794, 0.5864).

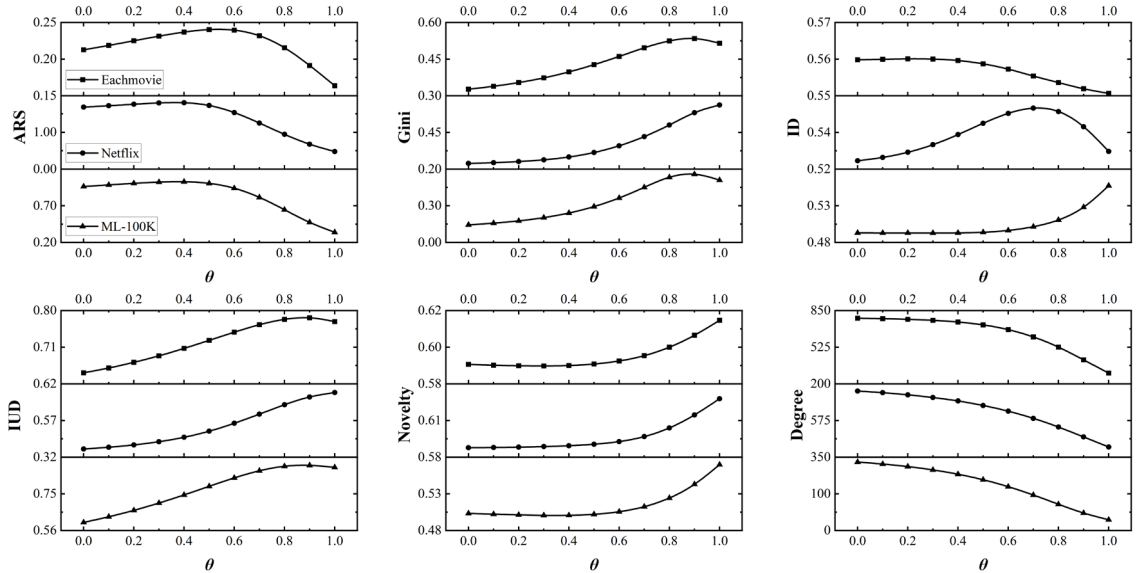

**Fig. 10.** Dependent variation of each metric with $\theta$

### 5.2.3 Distribution of recommended times (RQ4)

To confirm the effectiveness of the PIM+RA algorithm in balancing item recommendation counts, **Fig. 11** visually illustrates the comparison of recommendation counts distribution among each algorithm. It can be visually observed that PIM+RA has a relatively balanced number of recommendations for items of different popularity in the three datasets compared to the MD algorithm, with small items receiving slightly more recommendations than large items. The MD algorithm shows the feature that the number of recommendations first increases and then tends to converge as the item degree increases. The recommendation resources of IBCF are mainly concentrated on small items, but this also is a kind of unbalanced performance of recommendation, resulting in an extremely low accuracy. The recommendation times of SVD and UBCF show similar characteristics, i.e., the distribution of recommended times for non-popular items is more scattered, while the recommendation times for popular items tend to be higher.

PIM+RA makes the number of recommendations for each item closer by appropriately directing recommendation resources from large-degree items to small-degree items and passing item similarity, which is an important reason for its balance of accuracy, coverage, diversity, and novelty. The comprehensive performance improvement of the recommendation algorithm is more limited by only recommending large-degree items or small-degree items.

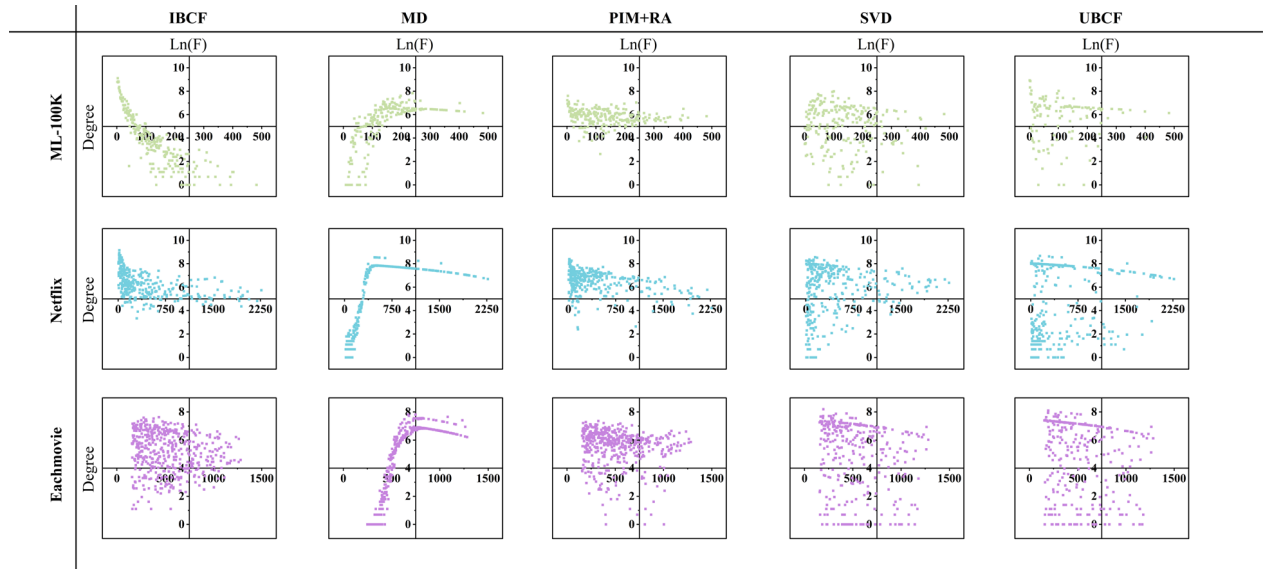

**Fig. 11**. Frequency of recommendations by each algorithm for items with different popularity. The horizontal axis indicates the popularity of the item, and the vertical axis indicates the logarithm of the recommendation frequency. The green, blue, and purple dots indicate the results of the ML-100K, Netflix, and Eachmovie datasets in order.

### 5.2.4 Effect of recommendation list length (RQ5)

In the above analysis, we show the evaluation results of each algorithm and metric for a fixed list length of 100. In order not to lose the generality of the results, the specifics of the four metrics related to the length of the recommendation list (Gini, ID, IUD, and Novelty) are discussed next for different lengths of the recommendation list and plotted in **Fig. 12**.

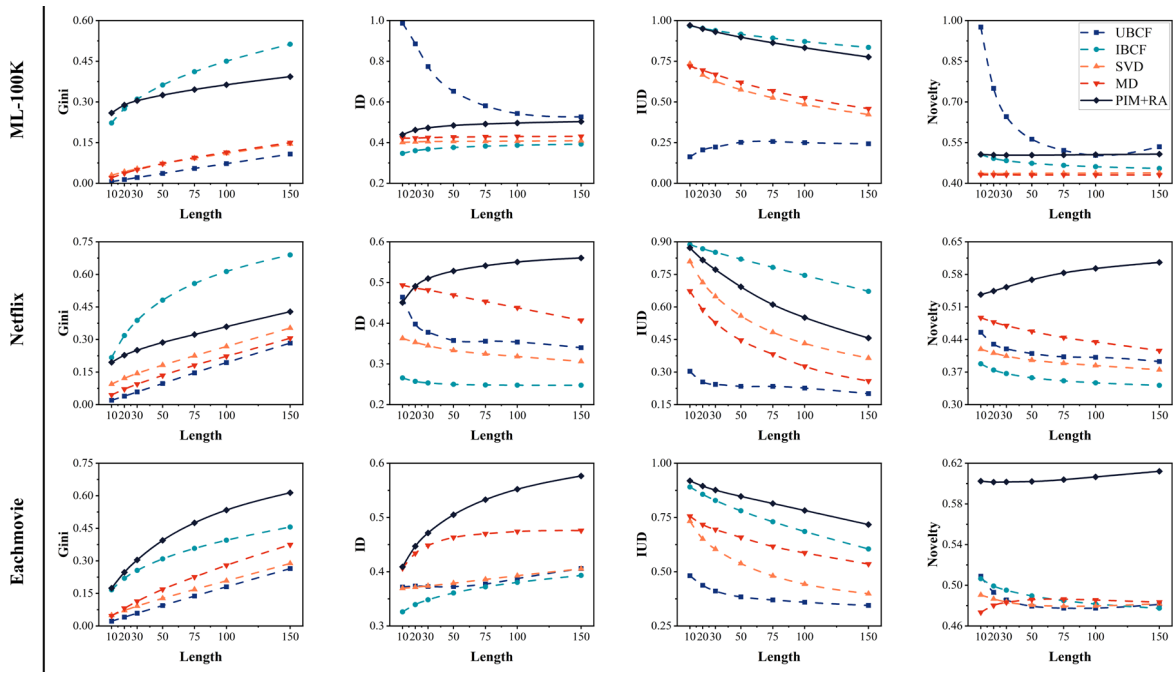

**Fig. 12.** Performance trends of five methods on three datasets with different recommendation list lengths as independent variables, the parameter setting of the PIM+RA method is the same as **Table 2**.

As shown in **Fig. 12**, the Gini trends of the five algorithms are similar, and they all show a monotonic increase with the increase of the recommendation list length, but the rate of change varies: PIM+RA and IBCF have a high to low rate of increase, while the other three algorithms have a more constant rate of increase. In the variation of ID, the other four algorithms all show a decrease in internal diversity as the length of the recommendation list increases, but PIM+RA maintains a monotonic increment and relatively outperforms most of the algorithms. It indicates that the PIM+RA algorithm can ensure the diversity of the top-ranked items within a certain range of recommendation list lengths and provide users with rich choices.

In the performance of IUD, the inter-user diversity decreases monotonically with increasing length of the recommendation list, except for ML-100K·UBCF scenario, and the growth rate decreases. PIM+RA and IBCF are better in all scenarios. As for novelty performance, PIM+RA is inferior to UBCF in the short recommendation list scenario of ML-100K, but the performance is flat when the recommendation list grows and is substantially better than other algorithms in the other two datasets. The novelty of PIM+RA tends to increase with the recommendation list length, which is different from the decreasing trend of other algorithms and shows that the PIM+RA algorithm is reasonable for the overall ranking.

Overall, the PIM+RA algorithm achieves relatively stable and excellent performance under different recommendation list lengths and different metrics.

## 6 Conclusion

This paper proposes the PIM+RA network resource allocation method, which integrates improved similarity to balance the accuracy of recommendation with other performances, while ensuring adequate exposure for recommendations of long-tail items. To achieve this, an improved bipartite network has been developed, supplemented with edge weights and self-connecting edges for resource allocation. A three-step resource allocation process, which accounts for user activity and item popularity, has been formulated. Additionally, the PIM+RA has the capability to adjust the recommendation list based on a parameter. The PIM similarity, grounded on the improved Pearson correlation coefficient, which comprehends the influence of co-rating-item, user activity, and item popularity on similarity calculation. It has also been incorporated into the bipartite network, resulting in accurate resource allocation.

The experimental results of the three datasets indicate that the PIM enhances recommendation accuracy in two collaborative filtering scenarios compared to PCC and CS, and results in a more uniform distribution of similarity. The PIM+RA attains the highest accuracy compared to four benchmark methods by selecting appropriate parameters and ensuring the stability and excellence of the four metrics, namely recommendation coverage, internal diversity, inter-user diversity, and novelty.

In the future, we consider automatically selecting the best parameter according to different recommendation scenarios and integrating expert user information (Duan et al. 2022), user trust matrix (Chen & Gao, 2018), and user social network (Yu et al. 2021), etc. for user-side messaging to further optimize the performance of the recommendation algorithm in all aspects.

## Declaration of interests

The authors declare that they have no known competing financial interests or personal relationships that could have appeared to influence the work reported in this paper.

**Acknowledgments**

This work received support from the Hunan Provincial Natural Science Foundation of China and Graduate Student Research and Innovation Fund of Central South University (Grant number: 2021JJ30031 and 1053320220429).

**Huiyu Li** is an M.S. degree candidate in Management Science and Engineering at the School of Business, Central South University, China. His research areas include recommender system and natural language processing.

**Pei Liang** is a Ph.D.degree candidate in Management Science and Engineering at School of Business, Central South University, China. Her primary research interests lie in decision-making theory and its application in medical big data analysis.

**Junhua Hu** received his Ph.D.degree in Management Engineeringfrom Chiba Institute of Technology, Japan, in 2004. He is currently a Professor in the Department of Management Science and Information Management, School of Business, Central South University, China. His current research interests include decision-making theory and its application in medical decision, risk management, and informationmanagement.